\documentclass[a4paper]{jpconf}
\usepackage{graphicx}
\usepackage{hyperref}
\usepackage{url}
\begin{document}
\title{The GMT detector alignment in the STAR experiment}

\author{N Ermakov (for the STAR Collaboration)}

\address{National Research Nuclear University MEPhI (Moscow Engineering Physics Institute), Kashirskoe highway 31, Moscow, 115409, Russia}
\ead{coffe92@gmail.com}

\begin{abstract}
The Solenoidal Tracker At RHIC (STAR) uses the Time Projection Chamber (TPC)
to perform tracking and particle identification.
In order to improve the corrections (such as space charge) and
monitor non-static distortions of the TPC, GEM-based chambers (GMT) were installed at eight
locations outside the TPC where they will provide optimal sensitivity to the distortions.
In order to reach this goal, the ionization clusters were measured by using the ADC signals
in each module. The positions of clusters and their deviations from track projections enabled alignment
of the GMT modules with respect to TPC to an accuracy $\sim200 \mu m$.
\end{abstract}

\section{Introduction}

The physics program of Relativistic Heavy Ion Collider (RHIC) aims to study the
nuclear matter at extreme conditions. The Solenoidal Tracker At RHIC (STAR)~\cite{star} allows
to measure properties of the matter created in the heavy--ion collisions by using
various detectors. The main detector of STAR is the Time Projection Chamber
(TPC)~\cite{tpc}. It allows to perform particle tracking and identification in
a wide momentum range. With increase of beam luminosities, it has become
crucial to reduce systematic uncertainties due to the TPC alignment and
distortion corrections (for instance, due to the space charge~\cite{space}).
In order to reach this goal eight GEM Chambers to Monitor the TPC
Tracking (GMT) were placed outside TPC at the Time-Of-Flight
(TOF)~\cite{tof} radius.
To utilize the GMT module's design resolution of $\sim$150$\mu$m in
both in-plane axes, we aim to align the modules with respect to
the TPC to a similar level.
Using the positions and deviations of the ionization clusters that were reconstructed
from the ADC signals one can perform the alignment procedure.

\section{Results and Discussions}
A charged particle that passes the GMT modules will produce primary ionization along its track.
The ionization electrons are registered by strips and pads
that correspond to the beam (V) and transverse to the beam (U) directions, respectively.
Using the information about the sizes of strips, pads and spaces between them allows to
reconstruct coordinates of the signal. Analog signal from each strip and pad is converted
to digital by Analog--to--Digital Converter (ADC). The ADC spectrum contains a
non--zero minimum value that corresponds to zero energy deposited, called the pedestal.
In the current analysis it was assumed that the first five events are pedestals.
Then, for each subsequent event, the calculated pedestal was subtracted from ADC signal.
The ADC's signal are not centered at zero outside of the cluster locations due to some low-level unresolved issues
with the method which have been deemed inconsequential for the purposes of using the devices for calibrations.
The TSpectrum package~\cite{tspect} was used to find peak channels
in the ADC signals (see red triangles on the figure 1) as a function of channel position.
If the peak with the maximum amplitude was 3$\sigma$ above the background then
it was accepted as a cluster. The clusters were fit by the Gaussian distribution.
In the case when no clusters were found, the ADC spectrum was added to the pedestal distribution.
Figure 1 shows an example of the ADC spectra for two found clusters in both V~(a)
and U~(b) directions within a single module.
During the work, it was found that the module 6 does not work.
\begin{figure}[h]
  \centering
  \includegraphics[width=0.6\textwidth] {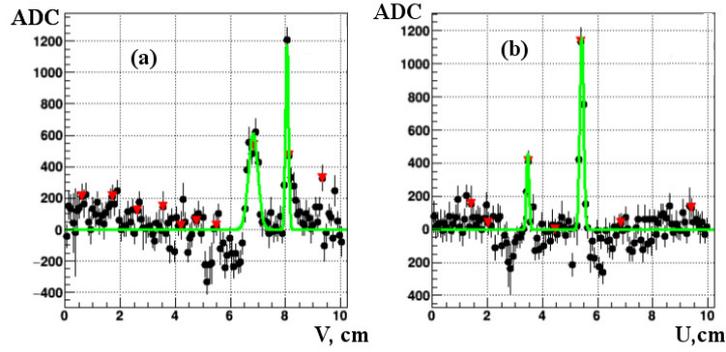}
  \caption{The example of the ionization cluster selection in the GMT module after the pedestal subtraction in the
    beam (a) and transverse (b) directions. The line represents fit of the clusters (see text).
  The red triangles correspond to peaks that were found by TSpectrum.}
\end{figure}
\begin{figure}[h]
  \centering
  \includegraphics[width=1.0\textwidth] {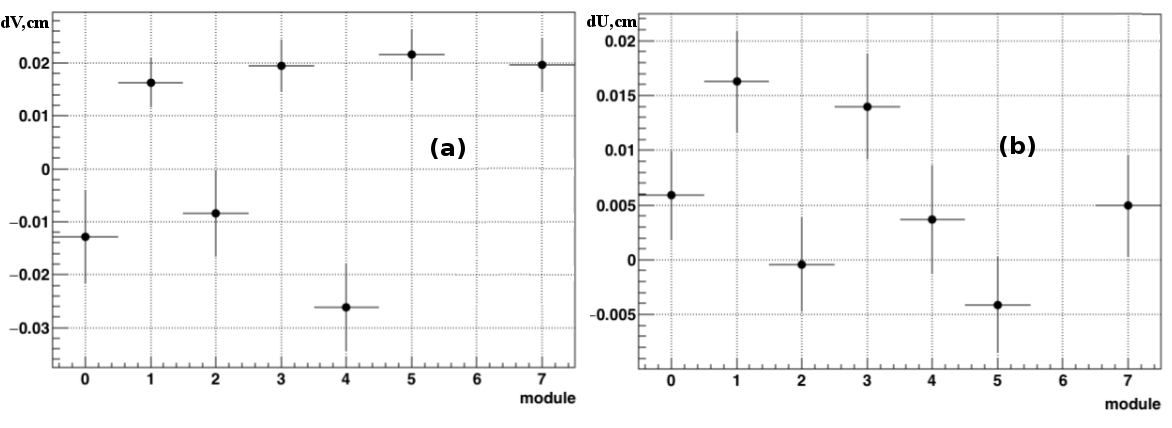}
  \caption{The alignment of the module in the beam (a) and transverse (b) directions with respect to TPC.}
\end{figure}
The clusters in both projections were matched by the amplitude
of the ADC signals.
In order to align the GMT modules with respect to TPC, the positions of clusters were
compared with projections of tracks from TPC.
Reasonable precision for the alignment may be achieved by using high statistics.
The method was tested on tracks from total 0.6 million Au+Au collisions at $\sqrt{s_{NN}}=14.5$~GeV
collected in 2014.
The $\sim$5$\times 10^{4}$ tracks that produced hits in GMT modules were selected
and allowed to align the GMT modules with respect to TPC with accuracy better than  $\sim200\mu$m.
Figure 2 shows the alignment of the GMT modules with respect to TPC in both U (a) and V (b) directions.
As a result of different statistics in each module the accuracy of the alignment depends on the module position.
In future with  higher statistics, it will be possible to improve
the alignment further and take into account other degrees
of freedom in the alignment,
such as rotations of the GMT modules.
\section{Conclusion}
The GMT modules will allow improved monitoring and
correction of TPC distortions.
In order to use the GMT modules for this purposes,
alignment of the GMT modules is necessary, and was performed by comparing
position of ionization clusters with the projections of TPC tracks.
The projections of TPC tracks were compared with cluster positions.
This allowed to align the GMT modules with respect to TPC with accuracy better than $\sim200\mu$m.
\section*{Acknowledgments}
This work was partially supported by MEPhI Academic Excellence Project (contract No. 02.a03.21.0005, 27.08.2013).
\section*{References}
\bibliographystyle{iopart-num}
\bibliography{Ermakov_GMT}

\providecommand{\newblock}{}
\begin{thebibliography}{1}
\expandafter\ifx\csname url\endcsname\relax
  \def\url#1{{\tt #1}}\fi
\expandafter\ifx\csname urlprefix\endcsname\relax\def\urlprefix{URL }\fi
\providecommand{\eprint}[2][]{\url{#2}}

\bibitem{star}
Ackermann K~H {\em et~al.\/} 2003 {\em Nucl. Instrum. Meth.\/} A {\bf 499} 624

\bibitem{tpc}
Anderson M {\em et~al.\/} 2003 {\em Nucl. Instrum. Meth.\/} A {\bf 499} 659

\bibitem{space}
{Van Buren} G {\em et~al.\/} 2006 {\em Nucl. Instrum. Meth.\/} A {\bf 566} 22

\bibitem{tof}
Llope W~J {\em et~al.\/} 2005 {\em Nucl. Instrum. Meth.\/} B {\bf 241} 306

\bibitem{tspect}
URL: \url{https://root.cern.ch/root/html530/TSpectrum.html}

\end{thebibliography}
\end{document}